\documentclass[11pt,a4paper]{article}
\pdfoutput=1
\usepackage{jstyle}
\usepackage{amsmath, amsfonts, amssymb}
\usepackage{graphicx, subfig, hyperref, color, verbatim}
\usepackage[dvipsnames]{xcolor}
\usepackage{float}
\usepackage{shuffle}
\newcommand{\bea}{\begin{equation}\begin{aligned}}
\newcommand{\eea}[1]{\label{#1}\end{aligned}\end{equation}}
\newcommand{\beq}{\begin{equation}}
\newcommand{\eeq}{\end{equation}}
\renewcommand \bar [1] {\overline{#1}}

\newcommand   \zb  {\overline{z}}
\newcommand   \mb  {\overline{m}}
\newcommand   \ub  {\overline{u}}

\newcommand   \cL  {\mathcal{L}}
\newcommand   \Res   {\mathop{\mathrm{Res}}}

\title{On the $AdS_3$ Virasoro-Shapiro Amplitude}
\author[a]{Luis F. Alday,}
\author[b]{Gaston Giribet,}
\author[c]{Tobias Hansen}
\affiliation[a]{Mathematical Institute, University of Oxford, Andrew Wiles Building, Radcliffe Observatory Quarter, Woodstock Road, Oxford, OX2 6GG, U.K.}
\affiliation[b]{Department of Physics, New York University. 726 Broadway, New York,
NY10003, USA.}
\affiliation[c]{Department of Mathematical Sciences, Durham University,
Stockton Road, Durham, DH1 3LE, U.K.}

\abstract{We consider tree-level scattering amplitudes for four string tachyons on $AdS_3 \times {\cal N}$ with pure NSNS fluxes. We show that in a small curvature expansion, properly defined, the amplitudes take the form of a genus zero integral given by the  Virasoro-Shapiro integrand with the extra insertion of single valued multiple polylogarithms. This is the same structure as the one found for the AdS Virasoro-Shapiro amplitude in higher dimensions.}

\begin{document}
\maketitle
\tableofcontents

\newpage

\section{Introduction}

In this paper we study tree level scattering amplitudes for string theory on AdS backgrounds. In flat space a standard textbook computation for the bosonic string leads to tree level amplitudes for any number of tachyons. The result for four points, denoted the Virasoro-Shapiro amplitude, is one of the most celebrated results in string theory. Superstring perturbation theory in flat space has been established decades ago \cite{Green:1981yb,Friedan:1985ge,DHoker:1988pdl}, when four point scattering amplitudes for massless states were computed. Both the standard Ramond-Neveu-Schwarz (RNS) as well as the pure spinor \cite{Berkovits:2000fe} worldsheet formalisms  can be used to compute supersymmetric amplitudes at tree level and by now compact expressions exist for any number of massless legs \cite{Mafra:2011nv}.  

In contrast, for curved space-times in the presence of RR-fluxes, the generic backgrounds of string theory, the computation of amplitudes, even at tree level, has been a major challenge. In this case the standard RNS formalism cannot be applied \cite{Metsaev:1998it} and alternative worldsheet formulations are not yet developed enough to compute, {\it e.g.} the analogue of the Virasoro-Shapiro amplitude. For curved space-times containing $AdS$ factors the AdS/CFT correspondence provides a definition of on-shell string scattering amplitudes in terms of correlators of local operators in the CFT at the boundary. This opens up the possibility of using (higher dimensional) CFT techniques where worldsheet techniques are not available. Combining CFT techniques with ideas from number theory progress has been made in the computation of the tree-level amplitude for the scattering of four gravitons in type IIB string theory on $AdS_5 \times S^5$, dual to the four-point correlator of stress-tensors in ${\cal N}=4$ SYM in the planar limit \cite{Alday:2022uxp,Alday:2022xwz,Alday:2023jdk,Alday:2023mvu}. More precisely, the amplitude can be computed in a large radius/small curvature expansion, and at each order it takes the form of a genus zero integral involving special functions introduced in \cite{Brown:2004ugm} and known as single valued multiple polylogarithms (SVMPLs).  This in turn implies that its low energy expansion contains only single valued Zeta values. This property has been established for tree-level massless closed superstring amplitudes in flat space \cite{Stieberger:2013wea,Stieberger:2014hba,Schlotterer:2018zce,Brown:2018omk,Brown:2019wna} and while still a conjecture for $AdS$, the proposal of \cite{Alday:2022uxp,Alday:2022xwz,Alday:2023jdk,Alday:2023mvu} passes several non-trivial tests. In particular, the first two curvature corrections to flat space were computed and the results were shown to reproduce all localisation \cite{Chester:2020dja} as well as integrability \cite{Gromov:2011de,Basso:2011rs,Gromov:2011bz} results, to the relevant order. 

A natural question is how to reproduce these results from a worldsheet perspective. In this endeavour it would be very interesting to have an example where the worldsheet is well established and we have full computational control of tree-level amplitudes. This is the case of string theory on $AdS_3 \times {\cal N}$ with pure NSNS fluxes. String theory in $AdS_3$ with pure NSNS fluxes is interesting mainly for two related reasons. Firstly, it provides a concrete example in which the theory can be solved on a curved background exactly; i.e., at finite values of $\alpha'$. Secondly, it represents a case of AdS/CFT where one can explore the correspondence exactly; in some cases, establishing the equivalence between bulk observables and boundary observables. The worldsheet $\sigma$-model describing the propagation of strings on $AdS_3$ with pure NSNS fluxes is given by a $SL(2,\mathbb{R})$ WZW model. This allows one to build up the spectrum of the theory and calculate correlation functions exactly \cite{Teschner:1997ft, Teschner:1999ug}; see also \cite{Giribet:2001ft, Maldacena:2001km}. In addition, this is an example where the dual CFT is two dimensional which enables one to employ the standard CFT$_2$ techniques. Early work on AdS$_3$/CFT$_2$ in the context of string theory allowed to establish the relationship between the worldsheet variables and the Virasoro symmetries of the dual CFT$_2$ theory \cite{Giveon:1998ns, deBoer:1998gyt, Kutasov:1999xu}; see also \cite{Giveon:2001up, Giveon:2005mi}. The spectrum of the theory was finally understood in \cite{Maldacena:2000hw}, and the analytic structure of correlation functions was studied in detail in \cite{Maldacena:2001km}. This enabled the comparison of various observables in the bulk and at the boundary \cite{Dabholkar:2007ey, Gaberdiel:2007vu, Giribet:2007wp}.  In recent years there have been several developments in the study of  $AdS_3$ strings in the context of holography, which allowed to understand  AdS$_3$/CFT$_2$ more precisely at some points in the moduli space \cite{Eberhardt:2017pty, Gaberdiel:2017oqg, Giribet:2018ada, Gaberdiel:2018rqv, Eberhardt:2019qcl, Eberhardt:2019ywk, Gaberdiel:2020ycd, Eberhardt:2020akk, Dei:2021yom, Balthazar:2021xeh, Eberhardt:2021vsx}.

In this paper we consider tree-level scattering amplitudes for string tachyons on $AdS_3 \times {\cal N}$ with pure NSNS fluxes. We develop an expansion around flat space and show that in this expansion amplitudes admit integral representations involving SVMPLs to all orders. This provides an example of an exactly solvable worldsheet theory, that leads to the structure found in \cite{Alday:2023jdk,Alday:2023mvu}. This paper is organised as follows. In section \ref{Sec:Amplitudes} we review strings on $AdS_3$ with pure NSNS fluxes, show how to compute tree level tachyonic amplitudes and give their general expression. In section \ref{Sec:expansion} we define the low curvature expansion - around flat space - and show that in this expansion the amplitudes are given by integrals involving SVMPLs, to all orders. We finish with some conclusions in \ref{Sec:disc}. The proof of $SL(2,\mathbb{C})$ invariance of the amplitudes and a brief description of SVMPLs are deferred to the appendices.

\section{String amplitudes on $AdS_3$}\label{Sec:Amplitudes}

\subsection{Strings on $AdS_3$}

We are interested in defining string theory on AdS$_3 \times \mathcal{N}$, where $\mathcal{N}$ is a compact internal manifold containing a $S^1$ factor. We first focus on the AdS$_3$ part of the geometry. The metric covering the Poincar\'e patch of AdS$_3$ is given by
\begin{equation}
ds^2=R^2\,\, \frac{dy^2 +dx^2-dt^2}{y^2},
\end{equation}
where $R$ is the radius of AdS$_3$. The boundary of the space is located at $y=0$. Defining 
$\gamma = x+t$, $\bar{\gamma}=x-t$, $y=e^{-\phi}$, the metric reads
\begin{equation}
ds^2=R^2\, \Big(d\phi^2 +e^{2\phi }d\gamma d\bar{\gamma}\Big),\label{metricon}
\end{equation}
where now the boundary is located at $\phi=\infty$. 

String theory supports the AdS$_3$ space provided one turns on NSNS and/or RR 2-form field(s). In this paper we consider the background with purely NSNS $B$-field flux, with the configuration being
\begin{equation}
B=R^2\, e^{2\phi }\,d\gamma \wedge d\bar{\gamma}\, .
\end{equation}
Evaluating the Polyakov action on this ansatz, we obtain the classical worldsheet  $\sigma $-model
\begin{equation}
S_{P}=\frac{R^2}{2\pi \alpha'}\int dz^2 \, \left(\partial \phi \bar{\partial}\phi +e^{2\phi }\partial \bar \gamma \bar\partial \gamma\right)\, .\label{S1}
\end{equation}
It is customary to define the dimensionless quantity
\begin{equation}
k=\frac{R^2}{\alpha '}\ , 
\end{equation}
whose square can be regarded as the 3-dimensional analogue to the t'Hooft coupling of AdS$_5$/CFT$_4$. From the point of view of the action (\ref{S1}) the semiclassical limit corresponds to large $k$. The action (\ref{S1}) for Lorentzian AdS$_3$ can be seen to be equivalent to the level-$k$ WZW action for the universal covering of the non-compact group $SL(2,\mathbb{R})$. Introducing two auxiliary fields, $\beta , \bar \beta$, and taking into account quantum corrections, the full action on AdS$_3\times S^1$ takes the form \cite{Giveon:1998ns}
\begin{equation}
S[\nu ]=\frac{1}{4\pi }\int d^2z \, \left(\partial \phi \bar{\partial}\phi -\sqrt{\frac{2}{k-2}} \mathcal{R}\,\phi+\beta \bar \partial \gamma + \bar \beta \partial \bar \gamma -4\pi \nu \beta \bar \beta e^{-\sqrt{\frac{2}{k-2}}\phi }+\partial X \bar{\partial}X\right)\, . \label{S2}
\end{equation}
Let us explain all its ingredients. $\mathcal{R}$ is the 2-dimensional Ricci scalar of the worldsheet. The dilaton term, which is linear in $\phi$, is generated by quantum corrections; this can be shown by carefully analysing the measure in the path integral \cite{Giveon:1998ns}. There is also a shift $k\to k-2$, which is a finite $\alpha '$ effect. In our conventions, we have canonically normalised the field $\phi $ by rescaling it by a factor $1/\sqrt{2k-4}$. Also, we have introduced the fields $\beta, \bar \beta$ which have no dynamics. By integrating them, in the large $k$ limit one recovers (\ref{S1}). $\nu$ can be interpreted as the inverse of the three-dimensional string coupling constant. It can be absorbed by shifting the zero mode of $\phi $, but it is convenient to keep it in order to control the coupling dependence. On the one hand, it is associated to the expectation value of the dilaton field, and, on the other hand, it enters in the string amplitudes as the genus-dependent KPZ scaling; see (\ref{esto}) below. Finally, the last term in (\ref{S2}) is that of a free scalar field $X$ that parameterises the $S^1$ part of the background. For convenience, $X$ has been canonically normalised as well.

The vertex operators that create Virasoro primaries in the worldsheet CFT are of the form
\begin{equation}
V_{j,m,\bar m, p}(z, \bar z) \, =\, \gamma^{m-j}(z) \,\bar \gamma^{\bar m-j}(\bar z) \, e^{-\sqrt{\frac{2}{k-2}}j\phi(z,\bar z)+i\sqrt{2}pX(z,\bar z)}\, .\label{vertex}
\end{equation}
These vertices correspond to non-excited, tachyon states. The label $j $ corresponds to the momentum of the string state in the radial direction $\phi $; more precisely, the radial momentum in string units is $j-\frac{1}{2} $. The angular momentum of the string state around the cylinder at the AdS$_3$ boundary is given by the difference $m-\bar m$, while the sum $m+\bar m$ corresponds to the kinetic energy of the state. $p=\bar p$ is the momentum along the $S^1$ direction in string units, assumed to be equal for left and right movers. The conformal dimension of the operator (\ref{vertex}) is given by
\begin{equation}
h=\bar h = \frac{j(1-j)}{k-2}+p^2\, .\label{dimension}
\end{equation}
The Virasoro constraint $h+\bar{h}=2$ yields the mass-shell condition. In (\ref{dimension}), $p\in {\mathbb{Z}}/{\sqrt{2} R_*}$, with $R_*$ being the radius of the $S^1$.

The complete spectrum in AdS$_3$ was constructed in \cite{Maldacena:2000hw}. This is organised in unitary representations of the universal covering of $SL(2,\mathbb{R})\times SL(2,\mathbb{R})$. Such representations are labelled by $j, m, \bar m$. The relevant representations for string theory are the highest- and lowest-state discrete series $\mathcal{D}^{\pm}_{j}$ (with $j\in \mathbb{R}_{<\frac{k-1}{2}}$, $\pm m=j+\mathbb{Z}_{\geq 0}$), together with the continuous principal series $\mathcal{C}^{\alpha }_j$ (with $j\in \frac 12 +i\mathbb{R}$, $\alpha \in \mathbb{R}$, $ m=\alpha +\mathbb{Z}$); see \cite{Maldacena:2000hw}. While the states belonging to discrete series $\mathcal{D}^{\pm}_{j}$ describe short strings confined in the bulk of AdS$_3$, the states of the continuous series $\mathcal{C}^{\alpha }_j$ describe long strings that can reach the boundary and thus define an S-matrix. In addition, the Hilbert space contains spectrally flowed representations, which are labelled by an extra quantum number $\omega \in \mathbb{Z}$ and correspond to winding string states. When $\omega \neq 0$ equation (\ref{dimension}) receives additional terms that depend on $\omega$, $m$ and $\bar m$; see \cite{Maldacena:2000hw}. Here, we are going to focus on the non-excited states of the spectral flow sector $\omega =0$, which are precisely those created by the operators (\ref{vertex}). 

In this paper we will consider tree level string amplitudes on $AdS_3\times S^1$. The tree level amplitude for $n$ external string states described by the vertex operators (\ref{vertex}), is given by the $n$-point correlator in the $SL(2,\mathbb{R})\times U(1)$ WZW model, integrated over the Riemann sphere:  
\begin{equation}
\mathcal{A}^{j_1,...,j_n;p_1,...p_n}_{m_1,...,m_n;\bar{m}_1, ... \bar{m}_n} = \int \prod_{i=1}^{n}d^2z_i\, \text{Vol$^{-1}$($PSL(2,\mathbb{C})$)}\, \left\langle \prod_{i=1}^{n} V_{j_i,m_i, \bar m_i, p_i}(z_i,\bar z_i) \right\rangle\,,
\end{equation}
where the expectation value is defined with respect to the action (\ref{S2}), namely
\begin{equation}
\left\langle \prod_{i=1}^{n} V_{j_i,m_i, \bar m_i, p_i}(z_i,\bar z_i) \right\rangle = \int \mathcal{D}\phi \mathcal{D}^2\gamma \mathcal{D}^2\beta \mathcal{D}X\, e^{-S[\nu ]} \prod_{i=1}^{n} V_{j_i,m_i, \bar m_i, p_i}(z_i,\bar z_i)\, .
\end{equation}
By integrating the zero mode $\phi_0=\phi-\tilde{\phi } $ one can prove that this expression yields \cite{Becker:1993at,Giribet:2001ft}
\bea
\left\langle \prod_{i=1}^{n} V_{j_i,m_i, \bar m_i, p_i}(z_i,\bar z_i) \right\rangle &{}= \nu^s\, \sqrt{k-2}\, \Gamma(-s) \int \prod_{r=1}^s\, \int \mathcal{D}\tilde \phi \mathcal{D}^2\gamma \mathcal{D}^2\beta \mathcal{D}X\, e^{-S[0]}  \\
&\times \, \prod_{i=1}^{n} V_{j_i,m_i, \bar m_i, p_i}(z_i,\bar z_i)\, \prod_{r=1}^s
\beta(u_r)\bar \beta (\bar u_r)e^{-\sqrt{\frac{2}{k-2}}\tilde{\phi}(u_r,\bar{u}_r)}
\eea{esto}
with
\begin{equation}
s=1-\sum_{i=1}^{n}j_i\, .\label{cond1}
\end{equation}
Here we have used the fact that we are interested in genus zero ($\text{g}=0$) amplitudes. For arbitrary genus, (\ref{cond1}) receives an additional contribution $-\text{g}$ on the right hand side. This, which can be easily seen from the coupling of the zero mode $\phi_0 $ to the Euler characteristic in the action (\ref{S2}), confirms the interpretation of $\nu$ as the inverse of the string coupling constant. In addition we have the following conservation rules for the amplitude to be non-vanishing
\begin{equation}
\sum_{i=1}^nm_i = 0 \, , \ \ \ \
\sum_{i=1}^n\bar m_i = 0 \, , \ \ \ \
\sum_{i=1}^np_i = 0 \, .\label{cond2}
\end{equation}
These follow from the integration over the other zero modes. 

Notice that on the right hand side of (\ref{esto}) the expectation value is defined in the theory with $\nu =0$. This reduces the computation of the $n$-string amplitudes to the computation of ($n+s$)-point correlators in a free theory consisting of two free scalars $X, \phi $ --the latter equipped with background charge-- and a ($1,0$)-dimension $\beta$-$\gamma$ ghost system. Indeed, that is the theory to which (\ref{S2}) reduces when $\nu=0$. In this free theory we simply have
\begin{equation}
\langle \tilde \phi(z) \tilde\phi (u)\rangle = -2\log |z-u|\, , \ \ \ \ 
\langle X(z) X(u)\rangle = -2\log |z-u|\, ,
\end{equation}
together with 
\begin{equation}
\langle \gamma (z) \beta (u)\rangle = -\frac{1}{(z-u)}\, , \ \ \ \ 
\langle \bar \gamma (\bar{z})\bar \beta  (\bar{u})\rangle =- \frac{1}{(\bar z-\bar u)}\,  .
\end{equation}
This implies 
\begin{equation}
\left\langle \prod_{i=1}^n e^{-\sqrt{\frac{2}{k-2}}j_i\tilde{\phi}(z_i,\bar z_i)}\, e^{-\sqrt{\frac{2}{k-2}}\tilde{\phi}(u,\bar u)}
\right\rangle \,=\, \prod_{i<i'}^n|z_i-z_{i'}|^{-\frac{4j_ij_{i'}}{k-2}}  \prod_{i=1}^n|z_i-u|^{-\frac{4j_i}{k-2}}\,  .
\end{equation}
and
\begin{equation}
\left\langle \prod_{i=1}^n\gamma^{m_i-j_i} (z_i)\, \beta (u)\right\rangle \,=\, 
\sum_{i=1}^n \frac{m_i-j_i}{u-z_i}  \,  , \ \ \ \
\left\langle \prod_{i=1}^n\bar \gamma^{m_i-j_i} (\bar z_i)\, \bar \beta (\bar u)\right\rangle \,=\, 
\sum_{i=1}^n \frac{\bar m_i-j_i}{\bar u-\bar z_i}  \, .
\end{equation}
Putting all together, tree level $n$-string amplitudes on AdS$_3\times S^1$ take the form \cite{Becker:1993at}
\begin{align}
{}&\mathcal{A}^{j_1,...,j_n;p_1,...,p_n}_{m_1,...,m_n;\bar{m}_1, ... ,\bar{m}_n} = \nu ^s\sqrt{{k-2}}\, {\Gamma(-s)}\, \int \prod_{i=1}^n d^2z_i\, \text{Vol$^{-1}$($PSL(2,\mathbb{C})$)}\, \, \prod_{i<i'} |z_i-z_{i'}|^{\frac{-4j_ij_{i'}}{k-2}+4p_{i}p_{i'}}\, \ \
\nonumber \\
&\times \int \prod_{r=1}^sd^2u_r\, 
\prod_{r=1}^{s} \prod_{i=1}^n |z_i-u_r|^{\frac{-4j_i}{k-2}}\prod_{r<r'}^{s} |u_r-u_{r'}|^{\frac{-4}{k-2}} \,
X^{-1}\frac{\partial^s X}{\partial{u_1}...\partial{u_s}}\, \bar X^{-1}\frac{\partial^s \bar X}{\partial{\bar u_1}...\partial{\bar u_s}}\,,
\label{Lamplitu}
\end{align}
with (\ref{cond1})-(\ref{cond2}) and with $X(z,u)=X(z_1,...,z_n;u_1,...,u_s)$ defined as follows
\bea
X(z_1, ...,z_n;u_1,...,u_s)&=&\prod_{r=1}^s\prod_{i=1}^{n}\,  (z_i-u_r)^{j_i-m_i}\, \prod_{l<t}^{s}(u_l-u_t)\,,\\
\bar X(\bar z_1, ..., \bar z_n;\bar u_1,...,\bar u_s)&=&\prod_{r=1}^s\prod_{i=1}^{n}\,  (\bar z_i-\bar u_r)^{j_i-\bar m_i}\, \prod_{l<t}^{s}(\bar u_l-\bar u_t)\, .
\eea{X_def}
In addition, we also have the mass-shell condition $h_i=\bar h_i =1$ for each external state, $i=1,2,...,n$. The expression (\ref{Lamplitu}) is manifestly crossing-symmetric. It is also invariant under $SL(2,\mathbb{C})$ transformations on the worldsheet, see appendix \ref{sl2c}. In the case of the 4-point amplitude ($n=4$), we can use this invariance to set $z_1=0$, $z_2=1$, $z_3=z$ and $z_4={\infty}$. This yields 
\begin{align}
\mathcal{A}^{j_1,...,j_4;p_1,...,p_4}_{m_1,...,m_4;\bar{m}_1, ... ,\bar{m}_4} ={}&\nu^s \sqrt{{k-2}} \, {\Gamma(-s)}\int d^2z\,  |z|^{-\frac{4j_1j_3}{k-2}+4p_1p_3}|1-z|^{-\frac{4j_2j_3}{k-2}+4p_2p_3}
 \\
& 
\int \prod_{r=1}^sd^2u_r\,
\, \left[\prod_{r<t}^{s} |u_r-u_t|^{-\frac{4}{k-2}}\prod_{r=1}^{s} \left( |u_r|^{-\frac{4j_1}{k-2}} 
|1-u_r|^{-\frac{4j_2}{k-2}}
|z-u_r|^{-\frac{4j_3}{k-2}}
\right)  \right.\nonumber
\\
&  \left. X^{-1}\frac{\partial^s X}{\partial {u_1}...\partial {u_s}}\, \bar X^{-1}\frac{\partial^s \bar X}{\partial{\bar u_1}...\partial{\bar u_s}}\,\right]\, \times \, \delta\left(\sum_{i=1}^4m_i\right) \, \delta\left(\sum_{i=1}^4\bar m_i\right) \, \delta\left(\sum_{i=1}^4p_i\right) \, \nonumber 
\end{align}
with $s=1-j_1-j_2-j_3-j_4$ and where now
\beq
X(z;u_1,...,u_s)=\prod_{r=1}^s\prod_{i=1}^{4}\,  u_r^{j_1-m_1}(1-u_r)^{j_2-m_2}(z-u_r)^{j_3-m_3}\, \prod_{l<t}^{s}(u_l-u_t)\, ,
\eeq
and analogously for its anti-holomorphic counterpart.

The generalisation to internal spaces containing multiple $S^1$ is straightforward. A prototypical example is $AdS_3 \times S^3 \times T^4$. In this case each $p_i$ is a vector and we simply replace products by inner products 
\begin{equation}
p_i p_j \to p_i \cdot p_j,
\end{equation}
with the momenta $p_i$ conserved along each circle. 

Before proceeding, let us make the following remark. The amplitude above involves $4+s$ integrals. The $s$ additional insertions can be thought of as the contributions from the background gravitons to the amplitude. These correspond to excited string states, with $j=1$, $p=0$ and level $N=1$, so that the on-shell condition 
\begin{equation}
\label{onshell}
h(j,p,N)=\frac{j(j-1)}{k-2}+p^2+N = 1\,,
\end{equation}
is indeed satisfied.

\subsection{Tree level amplitudes}

Let's focus on the case $n=4$ and write the amplitudes obtained above in the following way

\begin{equation}
\mathcal{A}_s(t_{13},t_{23})=  \int d^2z |z|^{2 t_{13}}|1-z|^{2 t_{23}} F^{j_1,j_2,j_3}_s(z) \,,
\label{AdS3VS}
\end{equation}
where we have ignored an overall prefactor, but the conservation rules, as well as the on-shell conditions are assumed. We have introduced the Mandelstam variables 
\begin{equation}
t_{13}=-\frac{2j_1j_3}{k-2}+2p_1 \cdot p_3,~~t_{23}= -\frac{2j_2j_3}{k-2}+2p_2 \cdot p_3 \,.
\end{equation}
The amplitude depends also on $j_i$ as well as on $m_i,\bar m_i$ but we have left this dependence implicit to ease the notation. $F_s(z)$ is an $s-$fold integral given by
\begin{equation}
F^{j_1,j_2,j_3}_s(z) = \int [du]\,
\, \prod_{r<t}^{s} |u_r-u_t|^{\frac{-4}{k-2}}\prod_{r=1}^{s} \left( |u_r|^{\frac{-4j_1}{k-2}} 
|1-u_r|^{\frac{-4j_2}{k-2}}
|z-u_r|^{\frac{-4j_3}{k-2}}
\right) \left|\frac{1}{X}\frac{\partial^s X}{\partial {u_1}...\partial {u_s}} \right|^2,
\end{equation}
where the integration measure is $[du]=\prod_{r=1}^sd^2u_r$. Note that (\ref{AdS3VS}) takes the form of the Virasoro-Shapiro amplitude in flat space, with the extra insertion of a function $F^{j_1,j_2,j_3}_s(z)$. The amplitude also has Bose symmetry under the exchange of any two operators. One can introduce an extra Mandelstam variable such that
\begin{equation}
\label{Mandelstanrels}
t_{13}+t_{23}+t_{43}= -2 + \frac{2 j_3 s}{k-2},~~~t_{43}=-\frac{2j_4j_3}{k-2}+2p_4 \cdot p_3\,.
\end{equation}
The following properties of the $s-$fold integral then imply crossing symmetry

\bea
F^{j_1,j_2,j_3}_s(1-z) &= F^{j_2,j_1,j_3}_s(z),\\
F^{j_1,j_2,j_3}_s \left(\frac{1}{z}\right) &= |z|^{\frac{4 j_3}{k-2}s} F^{j_4,j_2,j_3}_s(z),\\
F^{j_1,j_2,j_3}_s \left(\frac{z}{z-1}\right) &= |1-z|^{\frac{4 j_3}{k-2}s} F^{j_1,j_4,j_3}_s(z),
\eea{crossing}
where the change on the r.h.s, let's say $j_1 \to j_4$, also involves the corresponding change in $m_i,\bar m_i$.  

\subsection{Poles}

Let us study the poles of the amplitude in the $t_{13}$ plane. For the case $s=0$ we simply get the Virasoro-Shapiro amplitude/complex beta function

\begin{equation}
\mathcal{A}_{0}(t_{13},t_{23}) = \int d^2z |z|^{2 t_{13}}|1-z|^{2 t_{23}} = \frac{\Gamma(t_{13}+1)\Gamma(t_{23}+1)\Gamma(-1-t_{13}-t_{23})}{\Gamma(-t_{13})\Gamma(-t_{23})\Gamma(2+t_{13}+t_{23})}\,.
\end{equation}
Note that for $s=0$ the Mandelstam relations (\ref{Mandelstanrels}) reduce to
\begin{equation}
t_{13}+t_{23}+t_{43} = -2,
\end{equation}
so that the amplitude has indeed the correct crossing symmetries. In the $t_{13}$ plane it has a series of poles located at
\begin{equation}
t_{13} = -1,-2,-3,\cdots
\end{equation}
The location of these poles can be read off from the explicit answer, but also by considering the integral on a small disk around $z=0$, using polar coordinates. For general $s$ the location of the poles depends on the small $z$ expansion of $F^{j_1,j_2,j_3}_s(z)$. Consider first the case $s=1$
\begin{align}
F^{j_1,j_2,j_3}_{1}(z) ={}&\int d^2 u |u|^{-\frac{4j_1}{k-2}}|1-u|^{-\frac{4j_2}{k-2}}|z-u|^{-\frac{4j_3}{k-2}} \times \\
& \times \left(  \frac{j_1-m_1}{u} +  \frac{j_2-m_2}{u-1} + \frac{j_3-m_3}{u-z} \right) \left(  \frac{j_1-\bar m_1}{\bar u} +  \frac{j_2-\bar m_2}{\bar u-1} + \frac{j_3-\bar m_3}{\bar u-\bar z} \right) \nonumber.
\end{align}
This type of integrals was considered in \cite{Vanhove:2018elu,Vanhove:2020qtt}. They are single valued in the complex variable $z$ and satisfy second order differential equations in both $z$ and $\bar z$ (seen as independent variables). More precisely, introducing the notation
\begin{equation}
F^{a,b,c}_{\bar a,\bar b,\bar c}(z)=\int d^2 u \, u^a \bar u^{\bar a} (u-1)^b (\bar u-1)^{\bar b} (u-z)^c (\bar u -\bar z)^{\bar c},
\end{equation}
where we assume $a-\bar a \in \mathbb{Z}$, etc; one can show

\bea
\left(z(1-z)\partial_z^2 + \left( \left(a+b+2c \right)z -a-c \right) \partial_z -c(1+a+b+c) \right)  F^{a,b,c}_{\bar a,\bar b,\bar c}(z) = 0\,, \\
\left(\bar z(1-\bar z)\partial_{\bar z}^2 + \left( \left(\bar a+\bar b+2\bar c \right)\bar z -\bar a-\bar c \right) \partial_{\bar z} -\bar c(1+\bar a+\bar b+\bar c) \right)  F^{a,b,c}_{\bar a,\bar b,\bar c}(z) = 0\,.
\eea{hyp}
These equations can be solved in terms of hypergeometric functions. Introducing a basis of solutions
\begin{equation}
K^{a,b,c}_1(z) = {}_2F_1(-a-b-c-1,-c;-a-c;z ) ,\ \ K^{a,b,c}_2(z)= z^{a+c+1} \, {}_2F_1(a+1,-b;a+c+2;z),
\end{equation}
we can then write $F^{a,b,c}_{\bar a,\bar b,\bar c}(z)$ in terms of these. It turns out that the solution is diagonal, as it will be momentarily shown
\begin{equation}
F^{a,b,c}_{\bar a,\bar b,\bar c}(z) = \kappa_{11} K_1^{\bar a,\bar b,\bar c}(\bar z) K_1^{a,b,c}(z)+ \kappa_{22} K_2^{\bar a,\bar b,\bar c}(\bar z) K_2^{a,b,c}(z).
\end{equation}
This leads to the following two series in a small $z$ expansion
\begin{equation}
F^{a,b,c}_{\bar a,\bar b,\bar c}(z) \sim \left(\text{integer powers} \right)+z^{a+c+1} \bar z^{\bar a+\bar c+1} \times \left(\text{integer powers} \right)   \,, 
\end{equation}
where the integer powers are non-negative. The appearance of these two series can also be understood directly from the integral representation for $F^{a,b,c}_{\bar a,\bar b,\bar c}(z)$ and they arise from two distinct integration regions. This also allows to determine the constants $\kappa_{11},\kappa_{22}$ and show the diagonal form of the solution.  The first series, in integer powers, arises from the region of integration where $z$ is small, $|z| \ll |u|$. In this region we can expand $(u-z)^c (\bar u -\bar z)^{\bar c}$ in powers of $z,\bar z$. In particular this also implies
\begin{equation}
\kappa_{11} =\int d^2 u u^{a+c} \bar u^{\bar a+\bar c} (u-1)^b (\bar u-1)^{\bar b},
\end{equation}
which can be solved in terms of gamma functions for $a-\bar a,b-\bar b,c-\bar c$ integers, which is the case at hand. The second region corresponds to small $z$ but such that $|u| \sim |z|$. In this region we can change variables $u = z u'$ so that now $|u'|$ is not small. The integrand then reduces to
\begin{equation}
z^{a+c+1} \bar z^{\bar a+\bar c+1} \int d^2 u' u'^a \bar u'^{\bar a} (z u'-1)^b (\bar z \bar u'-1)^{\bar b} (u'-1)^c (\bar u' -1)^{\bar c}\,.
\end{equation}
We can now expand the integrand for small $z$, producing a series with integer powers times $z^{a+c+1} \bar z^{\bar a+\bar c+1}$. In particular, this also implies
\begin{equation}
\kappa_{22}=(-1)^{b+\bar b} \int d^2 u u^a \bar u^{\bar a} (u-1)^c (\bar u -1)^{\bar c}.
\end{equation}
Provided $a-\bar a,b-\bar b,c-\bar c$ are all integer, the final result for $F^{a,b,c}_{\bar a,\bar b,\bar c}(z)$ is single-valued in $z$, as expected from the fact that the original integrand is single valued. Going back to the series of poles, translating these results back to the case at hand, for $s=1$ the amplitude will have two series of poles, of the form 
\bea
&t_{13} = -1,-2,-3,\cdots\\
&t_{13}- 2\frac{j_1+j_3}{k-2} = -1,-2, -3,\cdots
\eea{poles2}
The same strategy can be applied for general $s$. In this case there will be $s+1$ distinct regions characterised by how many integration variables are small and how many are not. Assuming the first $n$ of them are small we can make the change of variables
\bea
&u_r = z u'_r,~~~&&\text{for } r=1,2,\cdots,n,  \\
&u_r = u'_r,~~~&&\text{for } r=n+1,\cdots,s, 
\eea{cov}
after which we can expand for small $z$ at the level of the integrand. This will lead to $s+1$ distinct series of poles, at locations
\begin{equation}
\label{poles}
t_{13} - 2\frac{j_1+j_3}{k-2} n -\frac{n(n-1)}{k-2}= -1, -2, -3,\cdots,~~~~\text{for }n=0,1,2,\cdots,s.
\end{equation}
These poles have the following interpretation. We have a tower of excited intermediate states at level $N$ and with $j=j_1+j_3+n$ and $p=p_1+p_3$, where $n=0,\cdots,s$, with $s$ the number of background gravitons contributing to the amplitude. The on-shell condition for such a state is then
\begin{equation}
h(j_1+j_3+n,p_1+p_3,N) = 1 \,,
\end{equation}
with the conformal dimension given by (\ref{onshell}). Using the on-shell conditions for the external particles $h(j_1,p_1,0)=h(j_3,p_3,0)=1$ this can be shown to be equivalent to
\begin{equation}
-\frac{2 j_1 j_3}{k-2}+2 p_1 \cdot p_3 - 2\frac{j_1+j_3}{k-2} n -\frac{n(n-1)}{k-2} = -1 -N,~~~N=0,1,2,\cdots
\end{equation}
which is exactly the location of the poles found above, see (\ref{poles}). 

\section{Expansion around flat space}\label{Sec:expansion}
In this section we will define an expansion where amplitudes display remarkable properties, very similar to those found in higher dimensions. We want to define a flat space limit and an expansion around it. We define it in such a way that the radius of $AdS_3$ becomes large, but so do the quantum numbers $j_i$, such that the Mandelstam variables $t_{ii'}$ remain fixed:

\begin{equation}
k = \frac{R^2}{\alpha'} \gg 1,~~~j_i \gg 1,~~~\text{with }  t_{i i'} = -\frac{2 j_i j_{i'}}{k-2}+2 p_i \cdot p_{i'} = \text{fixed}.
\end{equation}
In particular this implies $k \sim R^2$, $j_i \sim R$, $p_i \sim 1$. We will in addition assume that $s$ remains fixed in the limit, and consider cases with fixed $s=0,1,2,\cdots$. The integrand in the amplitudes (\ref{AdS3VS}) then splits into two factors. One takes the form of the usual Virasoro-Shapiro amplitude in flat space, and remains fixed in the limit. The other is the $s-$fold integral $F^{j_1,j_2,j_3}_s(z)$ whose expansion around flat space we now consider. 

\subsection{General prescription}

\subsubsection*{Case $s=1$}

The first non-trivial case corresponds to $s=1$ for which

\begin{equation}
F_{1}^{j_1,j_2,j_3}(z) = \sum_{a,b=1}^3 \int d^2 u |u|^{2 a_1}|1-u|^{2 a_2}|z-u|^{2 a_3} \frac{j_a-m_a}{u-z_a}\frac{j_b-\bar m_b}{\bar u-\bar z_b}\,,
\label{s1integral}
\end{equation}
where recall $z_1=0,z_2=1,z_3=z$ and we have introduced the ratios $a_i = \frac{-2 j_i}{k-2}$, which are small in the flat space limit. We could obtain the small $a_i$ expansion of the above expression from the result for $F_{1}^{j_1,j_2,j_3}(z)$ in terms of hypergeometric functions (see previous section). In the following, however, we will develop a more powerful method, that will apply to general $s$. When expanding in small $a_i$ we will find generic terms of the form
\begin{equation}
\int d^2 u   \frac{\log^p |u|^2 \log^q |1-u|^2 \log^r |z-u|^2}{(u-z_a)(\bar u-\bar z_b)}\,,
\end{equation}
for some non-negative integers $p,q,r$. Expressions of this form are single valued in $u$ and can be written as linear combinations of single valued polylogarithms (SVMPLs) ${\cal L}_w(u)$, where the words $w$ are formed by letters in the alphabet $\{0,1,z \}$ and have length/weight $p+q+r$. See appendix \ref{svmpl} for a brief account of SVMPLs. We are then led to integrate expressions of the form
\begin{equation}
\int d^2 u   \frac{{\cal L}_w(u)}{(u-z_a)(\bar u-\bar z_b)}.
\end{equation}
Similar integrations were analysed in \cite{Vanhove:2018elu,Vanhove:2020qtt}. In the present case, the integral can be performed as follows. First we use the defining property of SVMPLs to write

\begin{equation}
 \frac{{\cal L}_w(u)}{(u-z_a)(\bar u-\bar z_b)} = \partial_u \frac{{\cal L}_{z_a w}(u)}{\bar u-\bar z_b}\,,
\end{equation}
and then we use the following theorem \cite{Schnetz:2013hqa}

\begin{equation}
\int d^2u \partial_u f(u)  = \bar{\Res}_{u=\infty} f(u) -  \bar{\Res}_{u=z_b} f(u),~~~ f(u)=\frac{{\cal L}_{z_a w}(u)}{\bar u-z_b}\,.
\end{equation}
In particular, to order $p+q+r$ in the small $a_i$ expansion we expect SVMPLs of weight $p+q+r+1$. Furthermore, the r.h.s.\ of that expression can always be written in terms of SVMPLs in the variable $z$ with words from the alphabet $\{0,1\}$, see \cite{Panzer:2015ida,Vanhove:2018elu}. A generic term in the final expansion will then have the general form
\begin{equation}
\label{expansion1}
F_{1}^{j_1,j_2,j_3}(z)  = \cdots + a_1^p a_2^q a_3^r {\cal L}_W(z)+\cdots
\end{equation}
with $W$ a word formed with letters from the alphabet $\{0,1\}$ and weight $|W|=p+q+r+1$. Note that in doing this computation, we have assumed that we can expand the exponentials for small $a_i$ and then integrate term by term. This, however, is not quite true, and one has to be careful. More precisely for each of the nine contributions in  (\ref{s1integral}) there is a non-empty region of absolute convergence in the $(a_1,a_2,a_3)$ plane. For instance, the term proportional to $(j_1-m_1)(j_1-\bar m_1)$ in (\ref{s1integral}) converges absolutely in the region $\text{Re}(a_1)>0, \text{ Re}(a_2)>-1,\text{ Re}(a_3)>-1,\text{ Re}(a_1+a_2+a_3)<0$ which is non-empty. The integral is computed in this region and then the result extended to the point of interest, $a_i \to 0$, which is at the boundary of this region. This will lead to poles for small $a_i$ in the small $a_i$ expansion in (\ref{expansion1}). Indeed, note that the starting point of (\ref{expansion1}) will be a term with $|W|=0$, but this implies $p+q+r=-1$. In the next section these polar terms, as well as the whole expansion, will be computed carefully. 

\subsubsection*{General $s$}
For general $s$ we obtain

\begin{equation}
\frac{1}{X} \frac{\partial^s X}{ \partial u_1 \cdots \partial u_s} = \sum_{i_1,\cdots,i_s=1}^3 \frac{C_{i_1,\cdots,i_s}}{(u_1-z_{i_1})(u_2 -z_{i_2})\cdots(u_s-z_{i_s}) },
\end{equation}
where the sum contains $3^s$ terms and $C_{i_1,\cdots,i_s}$ are some constants. The $s-$fold integral then takes the form
\bea
F^{j_1,j_2,j_3}_s(z) ={}&\sum_{{\substack{i_1,\cdots,i_s=1 \\ \bar i_1,\cdots, \bar i_s=1}}}  \int [du]\,
\, \prod_{r<t}^{s} |u_r-u_t|^{2 \mu}\prod_{r=1}^{s} \left( |u_r|^{2 a_1} 
|1-u_r|^{2 a_2}
|z-u_r|^{2 a_3}
\right)\times\\
&\times \frac{C_{i_1,\cdots,i_s}}{(u_1-z_{i_1})(u_2 -z_{i_2})\cdots(u_s-z_{i_s}) }\frac{\bar C_{\bar i_1,\cdots,\bar i_s}}{(u_1-\bar z_{i_1})(u_2 -\bar z_{i_2})\cdots(u_s-\bar z_{i_s}) },
\eea{Fs}
where we have introduced the small ratio $\mu=-\frac{2}{k-2}$. Note that this ratio is parametrically smaller than $a_i$ in the flat space limit. We will perform the expansion in two steps: first we expand in powers of $a_1,a_2,a_3,\mu$, and then we insert their explicit $1/R$ dependence. Very much as for the case $s=1$ when expanding for small $a_i,\mu$ we will find generic terms of the form
\begin{equation}
 \int [du] \frac{{\cal L}_w(u_s)}{(u_1-z_{i_1})(u_2 -z_{i_2})\cdots(u_s-z_{i_s}) (u_1-\bar z_{i_1})(u_2 -\bar z_{i_2})\cdots(u_s-\bar z_{i_s})}\,,
\end{equation}
where we write the numerator in terms of SVMPLs in the variable $u_s$, with words formed by letters in the alphabet $\{0,1,z,u_1,\cdots,u_{s-1} \}$.\footnote{For example, for $s=2$ we would write $\log|u_1-u_2|^2= {\cal L}_{u_1}(u_2) + \log |u_1|^2 {\cal L}_e(u_2)$, with ${\cal L}_{u_1}(u_2)= \log \left|1-\frac{u_2}{u_1} \right|^2$ and ${\cal L}_e(u_2)=1$.}
We can now proceed and perform the integration over the variable $u_s$ using precisely the same method we described for $s=1$:
\begin{equation}
 \int du_s \frac{{\cal L}_w(u_s)}{(u_s-z_{i_s}) (\bar u_s-\bar z_{i_s})} = \overline{\Res}_{u_s=\infty} \frac{{\cal L}_{z_{i_s} w}(u_s)}{\bar u -\bar z_{i_s}}-\overline{\Res}_{u_s=z_{i_s}} \frac{{\cal L}_{z_{i_s} w}(u_s)}{\bar u -\bar z_{i_s}}\,.
\end{equation}
It turns out, see \cite{Panzer:2015ida}, that the right hand side can now be written as a linear combination of SVMPLs ${\cal L}_{w'}(u_{s-1})$ on the variable $u_{s-1}$ with words formed by letters in the alphabet $\{0,1,z,u_1,\cdots,u_{s-2} \}$. We are now left with the integral

\begin{equation}
 \int du_1 \cdots du_{s-1} \frac{{\cal L}_{w'}(u_{s-1})}{(u_1-z_{i_1})\cdots(u_{s-1}-z_{i_{s-1}}) (\bar u_1-\bar z_{i_1})\cdots(\bar u_{s-1}-\bar z_{i_{s-1}})}\,,
\end{equation}
but now we can repeat the same procedure to integrate over $u_{s-1}$ and so on, until we perform all the integrals. A generic term in the final expansion will then be of the form

\begin{equation}
F_{s}^{j_1,j_2,j_3}(z)  = \cdots + a_1^p a_2^q a_3^r \mu^t {\cal L}_W(z)+\cdots
\label{expansions}
\end{equation}
with $W$ a word formed with letters from the alphabet $\{0,1\}$ and weight $|W|=p+q+r+t+s$. Again, we have assumed that we can expand the exponentials for small $a_i,\mu$ and then integrate term by term. Let us now do this carefully.  

\subsection{Explicit results}

\subsubsection*{Case $s=1$}

Let us now tackle the integral for $s=1$, eq. (\ref{s1integral}), combining ideas from \cite{Vanhove:2018elu} with the method spelled out above. In order to do this integral in a small $a_i$ expansion, we would like to expand the exponentials into single valued logarithms, swap the sum with the integration and integrate term by term. However, the series is not absolutely convergent if either $a_i=0$ or $a_1+a_2+a_3=0$. To circumvent this issue we first consider the integral on the region $U_\epsilon = \mathbb{C} \setminus (B_0 (\epsilon) \cup B_1 (\epsilon) \cup B_z (\epsilon) \cup B_0 (\epsilon^{-1}) )$ where $B_x (r)$ is the ball centered at $x$ of radius $r$ and $\epsilon>0$. In this region, the exponential series is absolutely convergent, so we can swap it with the integration, and integrate term by term. We take $\epsilon \to 0$ at the end to obtain the original integral.

Recalling $\cL_{z^q} (u) = \log^q |1-\frac{u}{z}|^2 / q!$ for $z \neq 0$ and $\cL_{0^p} (u) = \log^p |u|^2 / p!$, together with the shuffle relations, see appendix \ref{svmpl}
\beq
\cL_w (z) \cL_{w'} (z) = \sum\limits_{W \in w \shuffle w'}  \cL_W (z)\,,
\eeq
we write
\begin{align}
F^{j_1,j_2,j_3}_{1,\epsilon}(z) \equiv{}&\int_{U_\epsilon} d^2 u |u|^{2a_1} |1-u|^{2a_2} |z-u|^{2a_3} 
\sum\limits_{a,b=1}^3 \frac{(j_a-m_a)(j_b-\mb_b)}{(u-z_a)(\ub-\zb_b)}\label{F1_exp}\\
={}& |z|^{2a_3}
\sum\limits_{a,b=1}^{3} (j_a-m_a)(j_b-\mb_b)
\int_{U_\epsilon} 
\hspace{-2pt}
d^2 u 
\hspace{-2pt}
\sum\limits_{p,q,r=0}^{\infty} a_1^p a_2^q a_3^r
\hspace{-6pt}
\sum\limits_{w\in 0^p \shuffle 1^q \shuffle z^r}
\hspace{-3pt}
\frac{\cL_w(u)}{(u-z_a)(\ub-\zb_b)}\,.
\nonumber
\end{align}
Following the idea of the previous subsection we can now identify
\beq
\partial_z f(z) = \frac{\cL_w(u)}{(u-z_a)(\ub-\zb_b)}
\quad \Leftrightarrow \quad
f(z) = \frac{\cL_{z_a w}(u)}{\ub-\zb_b}\,,
\eeq
and use Stokes theorem to obtain
\bea
F^{j_1,j_2,j_3}_{1,\epsilon}(z) ={}& |z|^{2a_3}
\sum\limits_{a,b=1}^{3} (j_a-m_a)(j_b-\mb_b)
\sum\limits_{p,q,r=0}^{\infty} a_1^p a_2^q a_3^r\\
&
\left(\oint_{\partial B_0 (\epsilon^{-1})} -\oint_{\partial B_0 (\epsilon)} - \oint_{\partial B_1 (\epsilon)} - \oint_{\partial B_z (\epsilon)}   \right)
\sum\limits_{w\in 0^p \shuffle 1^q \shuffle z^r}
\frac{\cL_{z_a w}(u)}{\ub-\zb_b} \frac{i d \bar{u}}{2 \pi}\,.
\eea{stokes}
The remaining contour integrals along the boundaries of $U_\epsilon$ receive two types of contributions. When integrating along $\partial B_{z_c}(\epsilon)$, the terms with $z_a = z_b = z_c$ have a logarithmic singularity at $u=z_c$. The second type of contribution are residues from the poles at $\ub = \zb_b$ and infinity. Let us first focus on the logarithmic singularity of the term with $z_a = z_b = 1$.
We observe that the two expressions
\beq
\sum\limits_{w\in 0^p \shuffle 1^q \shuffle z^r} \cL_{1 w}(u) \quad \text{and} \quad
\sum\limits_{w\in 0^p \shuffle 1^q \shuffle z^r} \frac{\cL_{1}(u) \cL_{w}(u)}{q+1}
= \sum\limits_{w\in 0^p \shuffle z^r}  \frac{\left(\log|1-u|^2 \right)^{q+1} \cL_{w}(u)}{(q+1)!}\,,
\label{sing}
\eeq
have the same logarithmic singularity at $u=1$, as they both contain the same SVMPLs whose label starts with 1.
Furthermore, the singularity is explicit in the last expression of \eqref{sing}.
Now we can use polar coordinates to do the integral.
\beq
-\oint_{\partial B_1 (\epsilon)} \sum\limits_{w\in 0^p \shuffle z^r} 
\frac{1}{(q+1)!} \frac{\left(\log|1-u|^2 \right)^{q+1} \cL_{w}(u)}{\ub - 1} \frac{i d \bar{u}}{2 \pi}
= -\sum\limits_{w\in 0^p \shuffle z^r} 
\frac{\left(\log \epsilon^2 \right)^{q+1}}{(q+1)!}  \cL_{w}(1) + O(\epsilon)\,.
\eeq
Plugging this back into \eqref{stokes} this contribution gives
\beq
-|z|^{2a_3}
\sum\limits_{p,q,r=0}^{\infty} a_1^p a_2^q a_3^r
\frac{\cL_0(1)^p}{p!}
\frac{\left(\log \epsilon^2 \right)^{q+1}}{(q+1)!} 
\frac{\cL_z(1)^r}{r!}
= \frac{1}{a_2}(1-\epsilon^{2 a_2} ) |1-z|^{2 a_3} \to \frac{1}{a_2} |1-z|^{2 a_3} \,,
\eeq
where in the last step we have sent $\epsilon \to 0$, assuming $a_2 >0$.\footnote{The contribution under consideration, proportional to $(j_2-m_2)(j_2-\bar m_2)$ in (\ref{s1integral}) converges absolutely in the region $Re(a_1)>-1,Re(a_2)>0,Re(a_3)>-1,Re(a_1+a_2+a_3)<0$ which is non-empty. The integral is computed in this region. Then we extend the result to the boundary, which is the limit of interest.}  The remaining logarithmic singularities can be treated similarly and we get a pole for each of the four contour integrals.
Besides this we get contributions from the residues, where we can now safely ignore the logarithmic singularities as they are already accounted for. The final result is
\begin{align}
F^{j_1,j_2,j_3}_{1}(z)
={}& \frac{(j_1-m_1)(j_1-\mb_1) |z|^{2 a_3}}{a_1}
+ \frac{(j_2-m_2)(j_2-\mb_2) |1-z|^{2 a_3}}{a_2}\nonumber\\
&+ \frac{(j_3-m_3)(j_3-\mb_3) |z|^{2 (a_1 + a_3)} |1-z|^{2 a_2}}{a_3}- \frac{\sum_{a=1}^3(j_a-m_a)\sum_{b=1}^3(j_b- \overline{m}_b) }{a_1+a_2+a_3}\nonumber\\
& + |z|^{2a_3}
\sum\limits_{a,b=1}^{3} (j_a-m_a)(j_b- \overline{m}_b)
\sum\limits_{p,q,r=0}^{\infty} a_1^p a_2^q a_3^r\nonumber\\
&\times\sum\limits_{w\in 0^p \shuffle 1^q \shuffle z^r}
\left( \overline{\Res}_{u=\infty} \frac{\cL_{z_a w} (u)}{\overline{u}- \overline{z}_b}
- \overline{\Res}_{u=z_b} \frac{\cL_{z_a w} (u)}{\overline{u}- \overline{z}_b}  \right)\,.
\label{F1_result}
\end{align}
The first two lines represent the polar contribution mentioned above. Note that introducing $a_4=-\frac{2 j_4}{k+2}$ such that $a_1+a_2+a_3+a_4=0$ and recalling that for $s=1$ we have $\sum_{i=1}^4 j_i =\sum_{i=1}^4 m_i =\sum_{i=1}^4 \bar m_i=0$, we can write this polar contribution in a completely symmetric fashion
\bea
F^{j_1,j_2,j_3}_{1,polar}(z)
={}& \frac{(j_1-m_1)(j_1-\mb_1) |z|^{2 a_3}}{a_1}
+ \frac{(j_2-m_2)(j_2-\mb_2) |1-z|^{2 a_3}}{a_2}\\
&+ \frac{(j_3-m_3)(j_3-\mb_3) |z|^{2 (a_1 + a_3)} |1-z|^{2 a_2}}{a_3}+ \frac{(j_4-m_4)(j_4-\bar m_4)}{a_4}.
\eea{F1_polar}
Note that each of these four polar contributions could have been computed by focusing on the relevant regions of integration. Going back to the full answer, to the first two orders in the small $a_i$ expansion $F^{j_1,j_2,j_3}_{1}(z)$ reads
\begin{align}
F^{j_1,j_2,j_3}_{1}(z) &=
\frac{(j_1-m_1)(j_1-\mb_1)}{a_1}
+ \frac{(j_2-m_2)(j_2-\mb_2)}{a_2}\nonumber\\
&+ \frac{(j_3-m_3)(j_3-\mb_3)}{a_3}
- \frac{\sum_{a}(j_a-m_a)\sum_{b}(j_b- \overline{m}_b) }{a_1+a_2+a_3}\label{F1_final}\\
&+ \frac{(a_1 (j_3-m_3)-a_3(j_1-m_1))(a_1 (j_3-\mb_3)-a_3(j_1-\mb_1))}{a_1 a_3} \log|z|^2\nonumber\\
&+\frac{(a_2 (j_3-m_3)-a_3(j_2-m_2))(a_2 (j_3-\mb_3)-a_3(j_2-\mb_2))}{a_2 a_3} \log|1-z|^2 + O(a_i)\,.
\nonumber
\end{align}
in perfect agreement with the general structure (\ref{expansion1}). In particular, note that the leading term in the expansion corresponds to weight zero SVMPLs (since it's independent of $z$) times rational functions (simple poles) of degree $-1$ in the $a_i$.

\subsubsection*{Case $s=2$}

For $s=2$ we have
\beq
\frac{1}{X} \frac{\partial^2 X}{ \partial u_2 \partial u_2} 
= \sum\limits_{i,k=1}^3 \frac{(j_i-m_i)(j_k-m_k + \delta_{i,k})}{(u_1-z_i)(u_2-z_k)}\,,
\eeq
where $z_1 = 0$, $z_2 = 1$, $z_3 = z$ and we have sent $z_4 \to \infty$. This leads to several contributions that can be treated in a similar way. As an example, let us focus on the diagonal term 
\begin{equation}
G_2^{a_1,a_2,a_3,\mu}(z)= \int d^2 u_1 d^2 u_2 \frac{|u_1|^{2a_1}|u_2|^{2a_1}}{|u_1|^2 |u_2|^2}|1-u_1|^{2a_2}|1-u_2|^{2a_2}|z-u_1|^{2a_3}|z-u_2|^{2a_3}|u_1-u_2|^{2\mu}.
\end{equation}
The region of absolute convergence of the integrals is $\text{Re}(a_1)>0,\text{ Re}(a_2)>-1,\text{ Re}(a_3)>-1,\text{ Re}(a_1+a_2+a_3+\mu)<0$. As we approach the boundary of this region (by taking $a_i,\mu$ small) we then expect two polar terms, one at $a_1=0$ and one at $a_1+a_2+a_3+\mu=0$. The first pole arises from the region of integration where at least one $u_i$ is very small. The second pole arises from the region where at least one $u_i$ is very large. These poles are related by symmetry. Indeed, by a change of variables in the integral above it can be shown that
\begin{equation}
G_2^{a_1,a_2,a_3,\mu}(1/z)= |z|^{-4 a_3} G_2^{-a_1-a_2-a_3-\mu,a_2,a_3,\mu}(z)\,.
\end{equation}
Let us compute the residue of the pole at $a_1=0$. To do this write an equivalent expression
\begin{equation}
G_2^{a_1,a_2,a_3,\mu}(z)=2  \hspace{-0.15in} \int \limits_{|u_1|< |u_2|}  \hspace{-0.15in} d^2 u_1 d^2 u_2 \frac{|u_1|^{2a_1}|u_2|^{2a_1}}{|u_1|^2 |u_2|^2}|1-u_1|^{2a_2}|1-u_2|^{2a_2}|z-u_1|^{2a_3}|z-u_2|^{2a_3}|u_1-u_2|^{2\mu},
\end{equation}
and now consider an expansion of the integrand around small $|u_1|$, which we will integrate term by term using polar coordinates (where the radius is integrated up to $|u_1|=|u_2|$). We obtain the following expansion
\beq
G_2^{a_1,a_2,a_3,\mu}(z)= \frac{|z|^{2a_3}}{a_1} G_1^{2a_1+\mu,a_2,a_3}(z)  + \cdots
\eeq
where $\cdots$ represent terms that are regular as $a_1 \to 0$ (for generic values of the other parameters) and we have defined
\beq
G_1^{a,b,c}(z) = \int d^2 u |u|^{2a-2}|1-u|^{2b}|z-u|^{2c}
\eeq
As an expansion around $a_1=0$ we then find
\beq
G_2^{a_1,a_2,a_3,\mu}(z)= \frac{|z|^{2a_3}}{a_1}  G_1^{\mu,a_2,a_3}(z) + \text{reg}.
\label{g2pole0}
\eeq
The pole at $a_1+a_2+a_3+\mu=0$ can then be computed by crossing symmetry. Note that the integral $G_1^{\mu,a_2,a_3}(z)$ was already met when discussing the $s=1$ case. It has the following expansion
\beq
G_1^{\mu,a_2,a_3}(z)= \frac{1}{\mu} - \frac{1}{\mu+a_2+a_3}+ \frac{a_3}{\mu} \log|z|^2+ \cdots.
\eeq
Plugging this into (\ref{g2pole0}) we find an expansion perfectly consistent with (\ref{expansions}). 

\subsection{Schematic structure of the expansions}
 
 The expansions of the $s-$fold integrals around flat space have the following schematic structure. Assume $a_i \sim \epsilon$ and $\mu \sim \epsilon$, all small and of the same order for now. Then we find
 
 \begin{equation}
F_s(z) = R^{(-s)}(a_i,\mu) \frac{{\cal L}_0(z)}{\epsilon^s} +R^{(-s+1)}(a_i,\mu) \frac{{\cal L}_1(z)}{\epsilon^{s-1}}+\cdots +R^{(q)}(a_i,\mu) \epsilon^{q} {\cal L}_{s+q}(z)+\cdots
 \end{equation}
 where ${\cal L}_k(z)$ denote combinations of SVMPLs of weight $k$ in $z$ with words from the alphabet $\{0,1\}$ and $R^{(q)}(a_i,\mu) $ are rational functions (different for each term in the combination) in $a_i,\mu$ of homogeneous weight $q$. This structure is perfectly consistent with the structure of poles  (\ref{poles}). Note that these poles collide in the flat space limit, leading to the usual poles of the Virasoro-Shapiro amplitude, and split as we take into account corrections. In an expansion, this splitting leads to higher order poles, which translate into logarithms at the level of $F_s(z)$. This structure is almost identical to the one found in \cite{Alday:2023jdk,Alday:2023mvu}! At each order in the curvature expansion we get the Virasoro-Shapiro amplitude in flat space, with the extra insertion of single-valued functions in exactly the same family. A difference is that here the weight/transcendentality jumps by one at each successive order in $1/R$, while in \cite{Alday:2023jdk,Alday:2023mvu} it jumps by three at each successive order in $1/R^2$. In the present case we are also studying amplitudes in a certain Mellin representation, since our vertex operators are related to vertex operators in the ``$x$-picture'' by the transform
 \begin{equation}
V_{j,m,\bar m}(z) = \int_{\mathbb{C}} d^2x V_j(z;x) x^{-j -m}\bar{x}^{-j -\bar m}\,,
 \end{equation}
where $x$ corresponds to the point on the boundary of Euclidean $AdS_3$ at which the operator is inserted. The Mellin and Borel transforms used in \cite{Alday:2023jdk,Alday:2023mvu}, however, are different. Another difference is the following. When writing our results in terms of $1/R$, we should note that $a_i \sim \frac{1}{R}$ while $\mu \sim \frac{1}{R^2}$. Hence, in a $1/R$ expansion terms of different weights will mix, and the discussion above applies to the highest weight. In  \cite{Alday:2023jdk,Alday:2023mvu} the weight is uniform, as the result of maximal supersymmetry. Another similarity is the appearance of rational functions in the variables $a_i$ of higher and higher degree. This corresponds to the polynomials in the variables $S,T$ appearing in  \cite{Alday:2023jdk,Alday:2023mvu}. A huge advantage of the present case, however, is that the answer is explicitly known and several properties and questions can be studied for finite radius.  Let us end with the following remark. Rewriting the amplitudes (\ref{AdS3VS})  as
\begin{equation}
\mathcal{A}_s(S,T)=  \int d^2z |z|^{-2S-2}|1-z|^{-2T-2} F_s(z) \,,
\end{equation}
our results imply that, to all orders around flat space, the low energy expansion around small $S,T$ will only contain single valued Zeta values.    
 
\section{Discussion and outlook}\label{Sec:disc}

In this paper we considered tree-level scattering amplitudes for four string tachyons on $AdS_3 \times {\cal N}$ with pure NSNS fluxes. The worldsheet is described by a $SL(2,\mathbb{R})$ WZW model, tree level amplitudes can be computed exactly and are given in terms of integral representations. We show that in a small curvature expansion, which we define, the amplitudes take the form of Virasoro-Shapiro integrals with the extra insertion of single valued multiple polylogarithms. This structure is almost identical to the one found in  \cite{Alday:2023jdk,Alday:2023mvu} with some differences, having to do with the somewhat different transforms used in the two problems. Some directions that would be interesting to explore are the following. 

Due to the full computational control of the worldsheet theory, strings on $AdS_3 \times {\cal N}$ with pure NSNS fluxes offer an ideal arena to study the ideas/structures of  \cite{Alday:2023jdk,Alday:2023mvu}, both in small curvature expansions as well as for finite radius. In this context it would be very interesting to study the fixed-angle and Regge high energy regimes, studied in a small curvature expansion in higher dimensions in \cite{Alday:2023pzu,Alday:2024xpq}. 

The results of this paper point to some universality for closed string amplitudes in curved backgrounds. In particular, single-valuedness plays an important role not only in flat space. The structure of the amplitudes in the present case imply that their low energy expansion will contain only single valued zeta values, to all orders. It would be very interesting to understand whether this is a universal feature of closed string amplitudes on curved backgrounds.  

A very interesting question in this general program is how to develop a worldsheet theory for strings on $AdS_5 \times S^5$ capable of reproducing tree level amplitudes in a small curvature expansion. The results of this paper represent a very neat example of how very similar structures arise from the $SL(2,\mathbb{R})$ WZW model. 

\acknowledgments
We would like to thank L. Eberhardt for useful discussions. The work of L.F.A.\ is partially supported by the STFC grant ST/T000864/1.
T.H.\ is supported by the STFC grant ST/X000591/1.

For the purpose of open access, the authors have applied a CC BY public copyright licence to any Author Accepted Manuscript (AAM) version arising from this submission.

\appendix

\section{$SL(2,\mathbb{C})$ invariance}\label{sl2c}

Consider the tree level $n-$string amplitude we obtained in the body of the paper, and let's rewrite it as
\begin{equation}
{\cal A}^{j,p}_{m,\bar m} = {\cal N} \int [dz][du] \prod_{i<i'}^n |z_i-z_{i'}|^{2 t_i \cdot t_{i'}} \prod_{r=1}^s \prod_{i=1}^n |z_i-u_r|^{-\frac{4 j_i}{k-2}} \prod_{r<r'}^s|u_r-u_{r'}|^{-\frac{4}{k-2}} \left| \frac{1}{X}\frac{\partial^s X}{\partial u_1 \cdots \partial u_s}\right|^2
\end{equation}
where $[dz][du] =\prod_{i=1}^n d^2 z_i  \prod_{r=1}^s d^2u_r$, $ {\cal N}$ is a $SL(2,\mathbb{C})$ invariant factor that will play no role in the following discussion and we have introduced 

\begin{equation}
t_i \cdot t_{i'} = -2\frac{j_i j_{i'}}{k-2} + 2 p_i \cdot p_{i'}.
\end{equation}
In this notation the on-shell conditions and momentum conservation take the form

\begin{equation}
t_i \cdot t_{i} =2 - \frac{2 j_i}{k-2},~~~\sum_{i=1}^n t_i =\left( i \sqrt{\frac{2}{k-2}}(1-s),0\right)\,,
\end{equation}
where we have introduced a vector notation
\begin{equation}
t_i =\left(i \sqrt{\frac{2}{k-2}} j_i,\sqrt{2} p_i \right).
\end{equation}
In particular note
\begin{equation}
\label{niceidentity}
2 t_i \cdot \sum_{i' \neq i}^n t_{i'} =2 t_i \cdot \left( \left( i \sqrt{\frac{2}{k-2}}(1-s),0\right) - t_i \right) = \frac{4 j_i s}{k-2}-4.
\end{equation}
Let us now make an $SL(2,\mathbb{C})$ transformation simultaneously in all the integration points $z_i,u_i$

\begin{equation}
z_i \to \frac{a z_i+b}{c z_i +d},~~~u_i \to \frac{a u_i+b}{c u_i +d},~~~~a d-b c=1.
\end{equation}
Under this transformation the integration measure picks up a factor
\begin{equation}
d^2 z_i \to \frac{d^2 z_i}{|c z_i+d|^4},~~~d^2 u_i \to \frac{d^2 u_i}{|c u_i+d|^4}\,,
\end{equation}
while distances behave as
\begin{equation}
|x-y|^2 \to \frac{|x-y|^2}{|c x+d|^2|c y+d|^2}\,,
\end{equation}
where $x,y$ are any of the integration variables. With these properties it is possible to show that under $SL(2,\mathbb{C})$ transformations
\begin{equation}
 \left| \frac{1}{X}\frac{\partial^s X}{\partial u_1 \cdots \partial u_s}\right|^2 \to \left(\prod_{r=1}^s |c u_r+d|^4 \right) \left| \frac{1}{X}\frac{\partial^s X}{\partial u_1 \cdots \partial u_s}\right|^2\,,
\end{equation}
the prefactor exactly cancels the factors picked by the integration measure $[du]$. We are then left with the extra factor

\bea
\text{extra} ={}& \left( \prod_{i=1}^n \frac{1}{|c z_i+d|^4} \right) \left(\prod_{i<i'}^n \frac{1}{| c z_i+d|^{2 t_i \cdot t_{i'}}|c z_{i'}+d|^{2 t_i \cdot t_{i'}}}\right) \times \\
&\times \left(\prod_{r=1}^s \prod_{i=1}^n |c z_i+d |^{\frac{4 j_i}{k-2}} |c u_r+d |^{\frac{4 j_i}{k-2}} \right) \left( \prod_{r<r'}^s|c u_r+d|^{\frac{4}{k-2}} |c u_{r'}+d|^{\frac{4}{k-2}} \right)\,.
\eea{extra}
By using $\sum_{i=1}^n j_i=1-s$ the second line can be simplified to 

\begin{equation}
\left( \prod_{i=1}^n |c z_i+d |^{\frac{4 j_i}{k-2} s} \prod_{r=1}^s|c u_r+d |^{\frac{4 (1-s)}{k-2}}\right) \prod_{r=1}^s |c u_r+d|^{\frac{4 (s-1)}{k-2}} = \prod_{i=1}^n |c z_i+d |^{\frac{4 j_i}{k-2} s} .
\end{equation}
Furthermore, using (\ref{niceidentity}) we can see
\begin{equation}
\prod_{i<i'}^n \frac{1}{| c z_i+d|^{2 t_i \cdot t_{i'}}|c z_{i'}+d|^{2 t_i \cdot t_{i'}}} = \prod_{i=1}^n \frac{1}{|c z_i +d|^{2 t_i \cdot \sum_{i' \neq i} t_{i'}}} =  \prod_{i=1}^n \frac{1}{|c z_i +d|^{\frac{4 j_i s}{k-2}-4}}\,,
\end{equation}
so that all factors precisely cancel and the amplitude is $SL(2,\mathbb{C})$ invariant. 

\section{Single valued multiple polylogarithms}\label{svmpl}
Let us start by defining multiple polylogarithms. These are holomorphic functions $L_w(z)$ labelled by a word $w$ formed of letters from an alphabet $\{ 0, \sigma_1,\sigma_2,\cdots \}$. For the empty word $e$ and the word with only $0's$ we have
\begin{equation}
L_e(z)=1,~~~L_{0^p}(z) = \frac{1}{p!} \log^pz,~~~p=1,2,\cdots
\end{equation}
For all other words we demand $L_w(z) \to 0$ as $z \to 0$, which fixes $L_w(z)$ recursively when supplemented by the differential relations
\begin{equation}
\frac{\partial}{\partial z} L_{\sigma_i w}(z) = \frac{ L_{w}(z)}{z-\sigma_i}.
\end{equation}
For instance, at weight one we obtain
\begin{equation}
L_{\sigma_i}(z) = \log \left(1-\frac{z}{\sigma_i} \right),~~~L_{0}(z) = \log z.
\end{equation}
As can be seen from these examples, multiple polylogarithms have branch cuts. It is possible to show, see \cite{Brown:2004ugm}, that there exists a unique family of single-valued functions ${\cal L}_w(z)$, denoted single valued multiple polylogarithms (SVMPLs) given by linear combinations of $L_{w'}(z)L_{w''}(\bar z)$, which satisfy the same differential relations    
\begin{equation}
\frac{\partial}{\partial z} {\cal L}_{\sigma_i w}(z) = \frac{ {\cal L}_{w}(z)}{z-\sigma_i}.
\end{equation}
such that ${\cal L}_{e}(z)=1$, ${\cal L}_{0^p}(z)= \frac{1}{p!} \log^p|z|^2$, for $p=1,2,\cdots$ and ${\cal L}_w(z) \to 0$ as $z \to 0$ for all other words. For words of length one we obtain 

\begin{equation}
{\cal L}_{\sigma_i}(z) = \log \left|1-\frac{z}{\sigma_i} \right|^2,~~~{\cal L}_{0}(z) = \log |z|^2.
\end{equation}
For words of length two and three there are various possibilities, and the resulting expressions in terms of classical polylogarithms are very complicated. From length four SVMPLs cannot be written in terms of classical polylogarithms. SVMPLs satisfy beautiful relations. In particular the shuffle identities 
\begin{equation}
{\cal L}_{w}(z) {\cal L}_{w'}(z) = \sum_{W \in w \shuffle w'}{\cal L}_{W}(z).
\end{equation}
We are often interested in evaluating SVMPLs at the special values $z=\sigma_j$, where $\sigma_j$ is one of the letters. ${\cal L}_{w}(\sigma_j)$ is then defined with the regularisation prescription that sets $\log 0=0$. Finally, the following theorem proven in  \cite{Panzer:2015ida,Vanhove:2018elu} is very useful. Namely
\begin{equation}
{\cal L}_w(\sigma_j) = \sum_{w'} c_{w'} {\cal L}(\sigma_i)\,,
\end{equation}
where on the r.h.s.\ we have a finite linear combination of SVMPLs in $\sigma_i$, now seen as the variable, with words from the alphabet $\{0,\sigma_1,\cdots \}/\sigma_i$ excluding $\sigma_i$.

\bibliography{refAdS3} 
\bibliographystyle{utphys}

\end{document}